\documentclass[sigconf]{acmart}

\AtBeginDocument{%
  \providecommand\BibTeX{{%
    \normalfont B\kern-0.5em{\scshape i\kern-0.25em b}\kern-0.8em\TeX}}}

\setcopyright{acmcopyright}
\copyrightyear{2022}
\acmYear{2022}
\acmDOI{XXXXXXX.XXXXXXX}

\acmConference[CIKM '22]{Make sure to enter the correct
  conference title from your rights confirmation email}{October 17--22,
  2022}{Atlanta, Ga}
\acmPrice{15.00}
\acmISBN{978-1-4503-XXXX-X/18/06}

\begin{document}

\title{Evaluating Dense Passage Retrieval using Transformers}

\author{Nima Sadri}
\email{nsadri@uwaterloo.ca}
\affiliation{%
  \institution{University of Waterloo}
  \country{Canada}
}

\renewcommand{\shortauthors}{Nima Sadri}

\begin{abstract}
Although representational retrieval models based on Transformers have been able to make major advances 
in the past few years, and despite the widely accepted conventions and best-practices for testing
such models, a \textit{standardized} evaluation framework for testing them has not been developed.
In this work, we formalize the best practices and conventions followed by researchers in the literature, paving the path for more standardized evaluations – and therefore more fair comparisons between the models. Our framework (1) embeds the documents and queries; (2) for each query-document pair, computes the relevance score based on the dot product of the document and query embedding; (3) uses the \texttt{dev} set of the MSMARCO dataset to evaluate the models; (4) uses the \texttt{trec\_eval} script to calculate MRR@100, which is the primary metric used to evaluate the models. Most importantly, we showcase the use of this framework by experimenting on some of the most well-known dense retrieval models. 
\end{abstract}


\ccsdesc[500]{Information systems~Information retrieval}
\ccsdesc[300]{Information systems~Language Models}
\ccsdesc[100]{Information systems~Retrieval models and ranking}

\keywords{transformers, neural networks, dense retrieval, MSMARCO}


\maketitle

\section{Introduction}
\begin{figure*}[h]
  \centering
  \includegraphics[width=\textwidth]{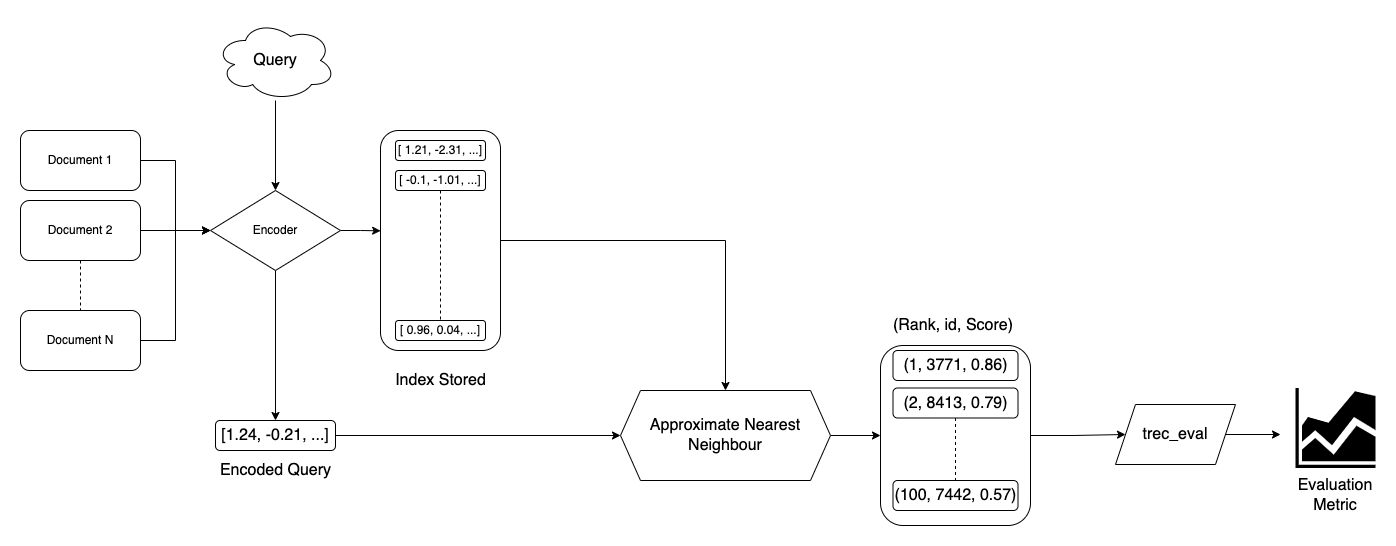}
  \caption{Our Proposed Framework for Standardizing the Evaluation Process}
\end{figure*}

In the past few years, pre-trained transformer-based models, such as BERT, T5, and the GPT family, have generated lots of excitement in NLP and Machine Learning. Most State-of-Art models in NLP for tasks such as translation, summarization, semantic analysis, and question-answering use some variant of a pre-trained transformer-based model. Even in information retrieval and text-ranking, we have seen the rise of new variants of such models: monoBERT, monoT5, DPR, to name a few. Since these models are resource-heavy, they increase the latency and cost of systems significantly, often for modest gains. Recent research has made great advances in decreasing the latency and costs of systems, while preserving their advantages, mostly by developing dense retrieval systems. Such systems first generate a representation for documents and queries independently, and then use a similarity function to score the document representations with respect to the query representation. Often times the document- and query-encoders are based on models such as BERT. Document embeddings can be pre-computed and stored on disk. When a query is received, its embedding can be calculated synchronously and high-ranking documents can be retrieved using the stored document-embedings. If the similarity function is simple (e.g. dot-product), the high-ranking documents can be approximately computed in an efficient manner, using Approximate Nearest Neighbour Search. 

There has been numerous methods proposed to generate document representations using transformer-based models. Although there has been extensive evaluations of such models on tasks such as sentence-similarity or semantic-analysis, a controlled comparison of these models for information retrieval is not present. Most research papers compare these models from the existing body of work, which may use different datasets, implementations, fine-tunning, and evaluation tools. However, as some time has passed, the research community has settled on standard implementations of these models, de facto datasets for ad-hoc retrieval, and standardized evaluation metrics and tools. We will put these pieces together, by evaluating the mostly used transformer-based representational methods on the widely used MSMARCO dataset using standard evaluation tools. We hope that this work provides a fair and objective evaluation of different representational retrieval models, and allows researchers to measure their models in a more standardized manner.

\section{Related Work}

\subsection{Sparse Embeddings}
Sparse document embeddings such as TF-IDF and BM25 (which is based on TF-IDF) have existed for 50 and 30 years, respectively. The embedings created by such models are sparse and high dimensional (dimension $\sim$ 50,000). They can be computed efficiently and can be interpreted fairly easily. However, they lack semantic meaning since they are based on exact-matches. For example, the embeddings for "That person teaches in college" and "She is a professor" would be orthogonal, despite the fact that both sentences have a similar meaning. This lack of semantic meaning is precisely what motivated lower-dimensional dense representations.

\subsection{Dense Embeddings}
To overcome the semantic mismatch present in exact-matching models such as BM25 and TF-IDF, researchers proposed low-dimensional (dimension $\sim$ 500) dense vectors which capture the semantic meaning of text. With the advent of Neural Networks, the word2vec approach \cite{word2vec} was proposed, which was one of the great success stories of self-supervision. In word2vec, a sliding window is passed through a large corpus of text (e.g. Wikipedia corpus) and an autoencoder-like neural network is trained to generate similar embedding for words that occur in a similar window. Similar approaches such as GloVe \cite{glove} were also proposed. These word embedings can be pooled to generate document embedings; this is the idea behind doc2vec \cite{doc2vec}. These models are categorized as self-supervised (also referred to as unsupervised) since they do not require human labelings. Embedings generated using word2vec, GloVe, and doc2vec encapsulate semantic meaning. For example, "That person teaches in college" and "She is a professor" would have embedings which are close to each other in the eucladian space. Therefore, dot-products can be used to measure the similarity of documents. 

The idea behind dense retrieval is as follows: given query $q$ and document $d$, use encoder $f$, $g$, and similarity metric $\phi$ to compute the relevance score as $\phi(f(d), g(q))$. In practice, $\phi$ might be dot-product and $f$ and $g$ might be transformer-based neural networks. Also, often times, $f = g$.

\subsection{BERT}
BERT \cite{bert} is a transformer-based language model that is trained in a self-supervised manner on a large corpus of text. Given sections of the corpus, BERT is trained on the following tasks:
\begin{enumerate}
\item Masked Language Modelling: random tokens are replaces with a special "[MASK]" token and the model is expected to predict what word was masked.
\item Next-Sentence Prediction: given 2 sentences, predict wheather the second sentence follows the first sentence in a corpus of text
\end{enumerate}
After training, BERT can generate embedding for a text that is equal to shorter in length to 512 tokens. Larger texts can be either truncated or split into several smaller texts to be able to passed through BERT.

\section{Evaluation}

\subsection{Datasets}
We use the MSMARCO dataset for evaluating our models. This is a large-scale real-life dataset consisting of queries given to the Bing search engine developed by Microsoft. It is important to note that most of the queries in the dataset have $1$ labled relevant document. Due to its size and real-life nature, the MSMARCO dataset has become one of the most popular datasets for ad-hoc information retrieval, especially when it comes to neural models.

More precisely, we use the \verb|dev| subset of the MSMARCO dataset. We perform the passage reranking task of MSMARCO. For each query $1,000$ candidate passages are identified, and the model's goal is to rerank them. This set contains $101,093$ queries and for each query $1,000$ candidate passages are identified. Since one passage may be relevant to multiple queries, in total there are $3,895,239$ unique passages that are used in the \verb|dev| reranking task.

Our models rerank the first-stage retrieval results provided by MSMARCO.

\begin{table}
  \caption{Examples of MSMARCO}
  \label{tab:msmarco}
  \begin{tabular}{p{0.35\linewidth} | p{0.6\linewidth}}
    \toprule
    Query (query id) & Relevant Document \\
    \midrule
    what is outer loop (1048554) & There are two lanes in the circle, one on the inside, one on the outside. The lane on the INSIDE is called the INNER LOOP, and the one on the OUTSIDE is called the OUTER LOOP. Inner loop - clockwise (curving right) (Closer to Washington D.C, ex. Driving from Fairfax to DC) Outer loop - Counter clockwise (curving left) (MD/VA Side, ex. \\
    \midrule
    what is outlook data file (1048730) & Outlook Data File (.pst) Outlook Data Files (.pst) contain your e-mail messages, calendars, contacts, tasks, and notes. You must use Outlook to work with the items in a .pst file. When you archive Outlook information, items are saved in a .pst files.\\
    \midrule
    what is peekaboo (1048554) &  Peekaboo definition, Also called bo-peep. a game played by or with very young children, typically in which one covers the face or hides and then suddenly uncovers the face or reappears, calling Peekaboo! See more. \\
    \midrule
    average number of lightning strikes per day (39449) &
Although many lightning flashes are simply cloud-to-cloud, there are as many as 9,000,000 reported lightning strikes that damage buildings, trees, and other objects every year. Worldwide, it is estimated that of an annual 1.4 billion lightning bolts, 25\% (more than 350 million) will strike the Earth's surface or objects on the surface. The vast majority of these strikes, however, occur in the tropics, and in unpopulated areas. 100 times per second; Lightning can strike over a thousand times in one storm. So, lightning strikes the earth over a million times a day. Globally, 8,640,000 lightning strikes per day. \\
    \midrule
    how much time does a lion spend sleeping (1095442) & Male lions spend 18 to 20 hours a day snoozing, while females get 15 to 18 hours of shuteye. The lionesses spend more time hunting and taking care of cubs, which is why they get slightly less sleep. \\
  \bottomrule
\end{tabular}
\end{table}

\subsection{Evaluation Metrics}
The main metric for MSMARCO is MRR@$100$. Therefore, we use this as our primary metric. To calculate this value use the standard \verb|trec_eval| script \footnote{\href{https://github.com/usnistgov/trec\_eval}{https://github.com/usnistgov/trec\_eval}}.

\section{Models}
\subsection{Sentence-BERT}
BERT \cite{bert} can be used to generate embedings for each of then input tokens. Sentence-BERT \cite{sbert} was one of the first works that used BERT to generate sentence embedding. Official implemention of Sentence-BERT is freely available \footnote{\href{https://www.sbert.net/}{https://www.sbert.net/}}. Different approaches was explored such as mean-pooling, max-pooling, and averaging the word embeddings. One of the most widely used techniques is using the CLS-embeding. The input sequence is prep-pended with a special classification token ("[CLS]") and the embedding of the CLS token is used as the representation of the sentence. This is sensible because BERT (and other transformer-based models) use the attention mechanism, which compute the attention operation for every pair of words. For example, say "[CLS] write with a pen" is passed through BERT. The attention between "[CLS]" and every other word (i.e. "write", "with", "a", and "pen") is computed. Moreover, Transformers have positional encodings, so the embeddings generated by the above example would be different if the order of the words is changed (e.g. "[CLS] with a pen write"). Therefore, the CLS-embedding encapsulates information about all the words in the sentence and their positions within the sentence. 

\verb|msmarco-bert-base-dot-v5| is based on the BERT-base variant of the co-condenser model \cite{cocondenser}. This model consists of stacks of Transformers. The [CLS] token is prepended to the input tokens and passed through the "early stacks" of Transformer blocks. The output is then passed through further Transformer stack (called the "late stacks"). Moreover, during the pre-training stage, the output of the late stacks are passed through another Transformer stack (called the "head stack"). The model is trained using a contrastive loss (hence then name Contrastive Condenser or CoCondenser. 

The \verb|msmarco-bert-base-dot-v5| model is trained on the MSMARCO dataset for semantic passage retrieval. 

The \verb|msmarco-distilbert-dot-v5| is a distilled version of BERT (Distilbert) \cite{distilbert} fine-tuned on the MSMARCO dataset for passage retrieval.

\verb|all-roberta-large-v1| is the RoBERTa model \cite{roberta}, which is a robustly optimized version of the large variant of BERT. The difference between RoBERTa and BERT is in the training process. More specifically, RoBERTa eliminates the next-sentence prediction task during pre-training and makes some hyperparamter changes to the training. In particular, RoBERTa uses much larger batch sizes and a larger learning rate. The experiments conducted on RoBERTa indicates that it is more robust to out-of-domain datasets when compared to the "vanilla" BERT model. The \verb|all-| prefix indicates that this model is trained on all the 1B sentence pairs for sentence embedding similarity; for convenience, we refer to models that are fine-tuned on this dataset collection as \verb|all-*| models.

\verb|multi-qa-mpnet-base-dot-v1| is based on the MPNet model \cite{mpnet}, which takes advantage of (1) the permuted language modeling for pre-training, which is based on XLNet; and (2) the full positional encoding information in BERT (in contrast to partial positional encoding in XLNet). It is fine-tuned on multiple passage retrieval tasks (not including msmarco).

\verb|all-mpnet-base-v2| is also based on the MPNet model, but it is fine-tuned for sentence embedding similarity on the \verb|all-*| dataset collection described previously.

\verb|all-MiniLM-L12-v2| is based on the MiniLM model \cite{minilm}, which is a light-weight and distiled version of BERT. This is achieved by training a student network to mimic the self-attention mechanism in BERT. The model is fine-tunned on the \verb|all-*| dataset collection. Similarly, \verb|msmarco-MiniLM-L6-cos-v2| is also based on the MiniLM model, but rather fine-tunned on the msmarco dataset for passage retrieval. 

\verb|all-distilroberta-v1| is based on the DistilBERT architecture, but the training is done by using the techniques that RoBERTa utilizes (described previously). This model is fine-tunned on the \verb|all-*| dataset collection.

\subsection{Contriever}
Similar to word2vec, the contriver \cite{cont} is trained in a self-supervised fashion by converting the corpus into different overlapping segments, using spans of length $n$. Then a contrastive loss is formulated to maximize agreement between these segments; that is, spans closer to each other should have more similar embedings when compared to segments farther away from each other. Using $f$ as the document/query encoder, $\phi$ as the similarity metric, $k_+$ as a positive document, $k_i$ as negative documents, and $q$ as query, the contrastive loss used is:
\begin{equation}
    \mathcal{L}(q, k_+) = \frac{exp(\phi(f(q), f(k_+)))}{\sum_{i=0}^{N^-} exp(\phi(f(q), f(k_i)))}
\end{equation}

The Contriever model uses a variety techniques for negative sampling to improve its performance; however, these details are beyond the scope of this work and can be found on the original paper \cite{cont}. The official implementation of the Contriever model is also available on GitHub \footnote{\href{https://github.com/facebookresearch/contriever}{https://github.com/facebookresearch/contriever}}.

\subsection{Topic Aware Sampling}
The TAS (Topic Aware Sampling) model \cite{tas}  selects training batches, consisting of positive and negative examples, in a novel fashion. The authors argue that when training batches are being chosen from a large dataset, choosing queries from different topics leads to random query-document interactions that contain little information gain. Their proposed model uses k-means clustering to separate the queries into k different sections, each pertaining to one "topic". Then to get a training batch on size $b$, $n$ topics ($n << k$) are randomly selected and for each of the chosen topics $\lfloor \frac{b}{n} \rfloor$ queries are selected from the respective cluster. Official implementation of the TAS model is available on GitHub \footnote{\href{https://github.com/sebastian-hofstaetter/tas-balanced-dense-retrieval}{https://github.com/sebastian-hofstaetter/tas-balanced-dense-retrieval}}
\begin{figure}[h]
  \centering
  \includegraphics[width=\linewidth]{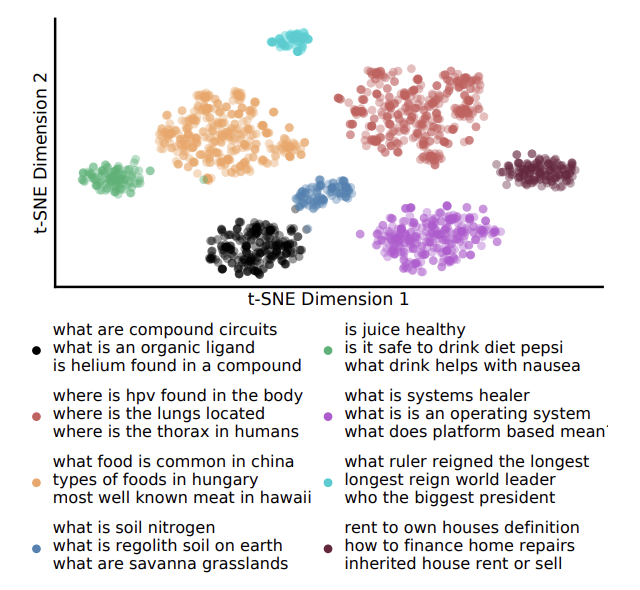}
  \caption{Random Examples of the Query Clusters Generated in TAS \cite{tas}}
\end{figure}

\section{Results}

\begin{table}
  \caption{Results on the MSMARCO Dataset}
  \label{tab:results}
  \begin{tabular}{lccc}
    \toprule
    Model & Similarity Metric & MRR@100 (\%)\\
    \midrule
    Top Leaderboard \footnote{TF-Ranking Ensemble of BERT, ROBERTA and ELECTRA }& - & 40.5 \\
    
    msmarco-bert-base-dot-v5 & dot-product & \textbf{37.56} \\
    msmarco-distilbert-dot-v5 & dot-product & \textbf{36.86} \\
    
    all-roberta-large-v1 & dot-product & 34.96 \\
    msmarco-distilbert-base-tas & dot-product & 34.53 \\
    msmarco-distilbert-base-tas-b & dot-product & 34.53 \\
    msmarco-TAS & dot-product & 34.51 \\
    multi-qa-mpnet-base-dot-v1 & dot-product & 34.38 \\
    msmarco-distilbert-cos-v5 & dot-product & 34.20 \\
    all-mpnet-base-v2 & dot-product & 33.76 \\
    all-MiniLM-L12-v2 & dot-product & 33.08 \\
    msmarco-MiniLM-L12-cos-v5 & dot-product & 32.74 \\
    all-distilroberta-v1 & dot-product & 32.45 \\
    msmarco-MiniLM-L6-cos-v5 & dot-product & 32.42 \\
    msmarco-contriver & dot-product & 18.08 \\

    

  \bottomrule
\end{tabular}
\end{table}

Recall that the official metric for MSMARCO is MRR@100, which is reported in Table \ref{tab:results} for the models we explored in this work. Moreover, we also included MRR@100 for the best model in the MSMARCO leaderboard (for the passage reranking task). This model uses an ensemble of BERT, ROBERTA and ELECTRA rerankers. \verb|msmarco-bert-base-dot-v5| and \verb|msmarco-distilbert-dot-v5| models performed comparable to the top leaderboard result, with only 2.5 points less percentage points of performance. Therefore, considering that these models are much simpler than the entries in the leaderboard, they still perform competetively. We note that, unlike the leaderboard entries, some of the our models were not fine-tuned on the MSMARCO dataset. In particular, the models that are not prefixed with \verb|msmarco-| have not been exposed to the MSMARCO dataset prior to this evaluation. For example, the \verb|all-roberta-large-v1| model has not been fine-tuned on the MSMARCO dataset, yet it only performs 5 percentage points worse than the top entry in the leaderboard. 

Despite its novel batch-selection technique, the \verb|msmarco-TAS| model does not seem to result in any improvements over the other models. Moreover, the \verb|msmarco-contriever| model's performance is surprisingly terrible. This is rather surprising, considering the fact that this model has been fine-tuned on the MSMARCO dataset and achieves competitive results in some retrieval tasks (including on the TriviaQA and NaturalQuestions datasets).

\section{Conclusion}

In this work, we formalized an standardized framework for evaluating representational retrieval models, which can be used to compare models in a fair and less biased fashion. We used best practices followed by the IR community and well-known tools (e.g. \verb|trec\_eval| script for evaluating the results). We then used our framework to evaluate a wide range of dense transformer-based models for ad-hoc information retrieval. We used the MSMARCO data for evaluating our models, which has become the defacto dataset for ad-hoc retrieval due to its large size and the fact that it was extracted from real search queries. The top 2 models in this work were fine-tunned on the MSMARCO dataset and achieved comparable results (only 2\% lower MRR@100) to the top MSMARCO learderboard entry, while being more simple and efficient. Notably, \verb|all-roberta-large-v1|, despite not having be fine-tunned on the MSMARCO dataset, was able to achieved comparable results as well (a 5\% drop in MRR@100 when compared to the top MSMARCO leaderboard entry). The IR literature is vast and we cannot possibly benchmark all the models (even when restricted to Transformer-based models) using our framework. However, we hope the community recognizes the importance of standardized evaluation methods and researchers in the future adopt our framework (or a variant thereof) to report results on a standardized evaluation process.

\begin{acks}
We thank Gordon Cormack for providing compute resources to support our experiments.
\end{acks}

\bibliographystyle{ACM-Reference-Format}
\bibliography{sample-base}


\begin{thebibliography}{12}


\ifx \showCODEN    \undefined \def \showCODEN     #1{\unskip}     \fi
\ifx \showDOI      \undefined \def \showDOI       #1{#1}\fi
\ifx \showISBNx    \undefined \def \showISBNx     #1{\unskip}     \fi
\ifx \showISBNxiii \undefined \def \showISBNxiii  #1{\unskip}     \fi
\ifx \showISSN     \undefined \def \showISSN      #1{\unskip}     \fi
\ifx \showLCCN     \undefined \def \showLCCN      #1{\unskip}     \fi
\ifx \shownote     \undefined \def \shownote      #1{#1}          \fi
\ifx \showarticletitle \undefined \def \showarticletitle #1{#1}   \fi
\ifx \showURL      \undefined \def \showURL       {\relax}        \fi
\providecommand\bibfield[2]{#2}
\providecommand\bibinfo[2]{#2}
\providecommand\natexlab[1]{#1}
\providecommand\showeprint[2][]{arXiv:#2}

\bibitem[Devlin et~al\mbox{.}(2019)]%
        {bert}
\bibfield{author}{\bibinfo{person}{Jacob Devlin}, \bibinfo{person}{Ming-Wei
  Chang}, \bibinfo{person}{Kenton Lee}, {and} \bibinfo{person}{Kristina
  Toutanova}.} \bibinfo{year}{2019}\natexlab{}.
\newblock \showarticletitle{BERT: Pre-training of Deep Bidirectional
  Transformers for Language Understanding}.
\newblock \bibinfo{journal}{\emph{ArXiv}}  \bibinfo{volume}{abs/1810.04805}
  (\bibinfo{year}{2019}).
\newblock


\bibitem[Gao and Callan(2021)]%
        {cocondenser}
\bibfield{author}{\bibinfo{person}{Luyu Gao} {and} \bibinfo{person}{Jamie
  Callan}.} \bibinfo{year}{2021}\natexlab{}.
\newblock \bibinfo{title}{Unsupervised Corpus Aware Language Model Pre-training
  for Dense Passage Retrieval}.
\newblock
\newblock
\urldef\tempurl%
\url{https://doi.org/10.48550/ARXIV.2108.05540}
\showDOI{\tempurl}


\bibitem[Hofstätter et~al\mbox{.}(2021)]%
        {tas}
\bibfield{author}{\bibinfo{person}{Sebastian Hofstätter},
  \bibinfo{person}{Sheng-Chieh Lin}, \bibinfo{person}{Jheng-Hong Yang},
  \bibinfo{person}{Jimmy Lin}, {and} \bibinfo{person}{Allan Hanbury}.}
  \bibinfo{year}{2021}\natexlab{}.
\newblock \showarticletitle{Efficiently Teaching an Effective Dense Retriever
  with Balanced Topic Aware Sampling}.
\newblock \bibinfo{journal}{\emph{Proceedings of the 44th International ACM
  SIGIR Conference on Research and Development in Information Retrieval}}
  (\bibinfo{date}{Jul} \bibinfo{year}{2021}).
\newblock
\urldef\tempurl%
\url{https://doi.org/10.1145/3404835.3462891}
\showDOI{\tempurl}


\bibitem[Izacard et~al\mbox{.}(2021)]%
        {cont}
\bibfield{author}{\bibinfo{person}{Gautier Izacard}, \bibinfo{person}{Mathilde
  Caron}, \bibinfo{person}{Lucas Hosseini}, \bibinfo{person}{Sebastian Riedel},
  \bibinfo{person}{Piotr Bojanowski}, \bibinfo{person}{Armand Joulin}, {and}
  \bibinfo{person}{Edouard Grave}.} \bibinfo{year}{2021}\natexlab{}.
\newblock \bibinfo{title}{Towards Unsupervised Dense Information Retrieval with
  Contrastive Learning}.
\newblock
\newblock
\urldef\tempurl%
\url{https://doi.org/10.48550/ARXIV.2112.09118}
\showDOI{\tempurl}


\bibitem[Le and Mikolov(2014)]%
        {doc2vec}
\bibfield{author}{\bibinfo{person}{Quoc~V. Le} {and}
  \bibinfo{person}{Tom{\'{a}}s Mikolov}.} \bibinfo{year}{2014}\natexlab{}.
\newblock \showarticletitle{Distributed Representations of Sentences and
  Documents}.
\newblock \bibinfo{journal}{\emph{CoRR}}  \bibinfo{volume}{abs/1405.4053}
  (\bibinfo{year}{2014}).
\newblock
\showeprint[arXiv]{1405.4053}
\urldef\tempurl%
\url{http://arxiv.org/abs/1405.4053}
\showURL{%
\tempurl}


\bibitem[Liu et~al\mbox{.}(2019)]%
        {roberta}
\bibfield{author}{\bibinfo{person}{Yinhan Liu}, \bibinfo{person}{Myle Ott},
  \bibinfo{person}{Naman Goyal}, \bibinfo{person}{Jingfei Du},
  \bibinfo{person}{Mandar Joshi}, \bibinfo{person}{Danqi Chen},
  \bibinfo{person}{Omer Levy}, \bibinfo{person}{Mike Lewis},
  \bibinfo{person}{Luke Zettlemoyer}, {and} \bibinfo{person}{Veselin
  Stoyanov}.} \bibinfo{year}{2019}\natexlab{}.
\newblock \bibinfo{title}{RoBERTa: A Robustly Optimized BERT Pretraining
  Approach}.
\newblock
\newblock
\urldef\tempurl%
\url{https://doi.org/10.48550/ARXIV.1907.11692}
\showDOI{\tempurl}


\bibitem[Mikolov et~al\mbox{.}(2013)]%
        {word2vec}
\bibfield{author}{\bibinfo{person}{Tomas Mikolov}, \bibinfo{person}{Kai Chen},
  \bibinfo{person}{Gregory~S. Corrado}, {and} \bibinfo{person}{Jeffrey Dean}.}
  \bibinfo{year}{2013}\natexlab{}.
\newblock \showarticletitle{Efficient Estimation of Word Representations in
  Vector Space}. In \bibinfo{booktitle}{\emph{ICLR}}.
\newblock


\bibitem[Pennington et~al\mbox{.}(2014)]%
        {glove}
\bibfield{author}{\bibinfo{person}{Jeffrey Pennington},
  \bibinfo{person}{Richard Socher}, {and} \bibinfo{person}{Christopher
  Manning}.} \bibinfo{year}{2014}\natexlab{}.
\newblock \showarticletitle{{G}lo{V}e: Global Vectors for Word Representation}.
  In \bibinfo{booktitle}{\emph{Proceedings of the 2014 Conference on Empirical
  Methods in Natural Language Processing ({EMNLP})}}.
  \bibinfo{publisher}{Association for Computational Linguistics},
  \bibinfo{address}{Doha, Qatar}, \bibinfo{pages}{1532--1543}.
\newblock
\urldef\tempurl%
\url{https://doi.org/10.3115/v1/D14-1162}
\showDOI{\tempurl}


\bibitem[Reimers and Gurevych(2019)]%
        {sbert}
\bibfield{author}{\bibinfo{person}{Nils Reimers} {and} \bibinfo{person}{Iryna
  Gurevych}.} \bibinfo{year}{2019}\natexlab{}.
\newblock \showarticletitle{Sentence-BERT: Sentence Embeddings using Siamese
  BERT-Networks}.
\newblock \bibinfo{journal}{\emph{Proceedings of the 2019 Conference on
  Empirical Methods in Natural Language Processing and the 9th International
  Joint Conference on Natural Language Processing (EMNLP-IJCNLP)}}
  (\bibinfo{year}{2019}).
\newblock
\urldef\tempurl%
\url{https://doi.org/10.18653/v1/d19-1410}
\showDOI{\tempurl}


\bibitem[Sanh et~al\mbox{.}(2019)]%
        {distilbert}
\bibfield{author}{\bibinfo{person}{Victor Sanh}, \bibinfo{person}{Lysandre
  Debut}, \bibinfo{person}{Julien Chaumond}, {and} \bibinfo{person}{Thomas
  Wolf}.} \bibinfo{year}{2019}\natexlab{}.
\newblock \bibinfo{title}{DistilBERT, a distilled version of BERT: smaller,
  faster, cheaper and lighter}.
\newblock
\newblock
\urldef\tempurl%
\url{https://doi.org/10.48550/ARXIV.1910.01108}
\showDOI{\tempurl}


\bibitem[Song et~al\mbox{.}(2020)]%
        {mpnet}
\bibfield{author}{\bibinfo{person}{Kaitao Song}, \bibinfo{person}{Xu Tan},
  \bibinfo{person}{Tao Qin}, \bibinfo{person}{Jianfeng Lu}, {and}
  \bibinfo{person}{Tie-Yan Liu}.} \bibinfo{year}{2020}\natexlab{}.
\newblock \bibinfo{title}{MPNet: Masked and Permuted Pre-training for Language
  Understanding}.
\newblock
\newblock
\urldef\tempurl%
\url{https://doi.org/10.48550/ARXIV.2004.09297}
\showDOI{\tempurl}


\bibitem[Wang et~al\mbox{.}(2020)]%
        {minilm}
\bibfield{author}{\bibinfo{person}{Wenhui Wang}, \bibinfo{person}{Furu Wei},
  \bibinfo{person}{Li Dong}, \bibinfo{person}{Hangbo Bao}, \bibinfo{person}{Nan
  Yang}, {and} \bibinfo{person}{Ming Zhou}.} \bibinfo{year}{2020}\natexlab{}.
\newblock \bibinfo{title}{MiniLM: Deep Self-Attention Distillation for
  Task-Agnostic Compression of Pre-Trained Transformers}.
\newblock
\newblock
\urldef\tempurl%
\url{https://doi.org/10.48550/ARXIV.2002.10957}
\showDOI{\tempurl}


\end{thebibliography}

\end{document}